# A Low Cost Two-Tier Architecture Model For High Availability Clusters Application Load Balancing

A B M Moniruzzaman, *StudentMember, IEEE*
*Department of Computer Science and Engineering*
*Daffodil International University*
*Dhaka, Bangladesh*
*abm.moniruzzaman.bd@ieee.org*

*Syed Akther Hossain, Member, IEEE & ACM*
*Department of Computer Science and Engineering*
*Daffodil International University*
*Dhaka, Bangladesh*
*aktarhossain@daffodilvarsity.edu.bd*

***Abstract*— This article proposes a design and implementation of a low cost two-tier architecture model for high availability cluster combined with load-balancing and shared storage technology to achieve desired scale of three-tier architecture for application load balancing e.g. web servers. The research work proposes a design that physically omits Network File System (NFS) server nodes and implements NFS server functionalities within the cluster nodes, through Red Hat Cluster Suite (RHCS) with High Availability (HA) proxy load balancing technologies. In order to achieve a low-cost implementation in terms of investment in hardware and computing solutions, the proposed architecture will be beneficial. This system intends to provide steady service despite any system components fails due to uncertainly such as network system, storage and applications.**

## I. Introduction

High-availability clusters provide continuous availability of services by eliminating single points of failure. Node failures in a high-availability cluster are not visible from clients outside the cluster. (High-availability clusters are sometimes referred to as failover clusters.) Red Hat Cluster Suite [1] provides high-availability clustering through its High-availability Service Management component.

Load-balancing clusters dispatch network service requests to multiple cluster nodes to balance the request load among the cluster nodes. Load balancing provides cost-effective scalability because you can match the number of nodes according to load requirements. If a node in a load-balancing cluster becomes inoperative, the load-balancing software detects the failure and redirects requests to other cluster nodes. Node failures in a load-balancing cluster are not visible from clients outside the cluster. Red Hat Cluster Suite provides load-balancing through LVS (Linux Virtual Server) [2].

This article focuses on how to build a two-tier architecture model combined with load-balancing technology and shared storage technology to achieve full facilities of the three-tier architecture for application load balancing e.g. web servers Cluster. This system can overcome in the cases of node failover, network failover, storage limitation and distributions load as like as the all the facilities of three-tier architecture model for high availability cluster for application load balancing. Web Sever clusters have gained much attention and have become increasingly popular for handling requests because of their unique design [3]. When handling large amounts of complex data, load-balancing is a crucial necessity [3].We use for application load balancing for web server cluster as a prototype for implementing this system.

In this research work, we propose a design and implementation of a low cost two-tier architecture model for high availability cluster combined with load-balancing and shared storage technology to achieve desired scale of three-tier architecture for application load balancing e.g. web servers. The paper is organized as follows: Section II discusses Tradition three-tier architecture model for application load balancing; Section III explores literature review of related works and motivation of this article; Section IV describes proposed design of model and implementation details of a low cost two-tier architecture model for high availability cluster combined with load-balancing and shared storage technology; Section V demonstrates performance test of model implementation; and finally Section VI. summarizes the paper work.

## II. Three-Tier Architecture Model for Application Load Balancing

The three-tier architecture consists of Load Balancer, which is the front-end machine of the whole cluster systems, and balances requests from clients among a set of servers, so that the clients consider that all the services is from a single IP address. Server Cluster, which is a set of servers running actual network services, such as Web, Mail, FTP, DNS and Media service. Shared Storage, which provides a shared storage space for the servers, so that it is easy for the servers to have the same contents and provide the same services [4].

For scalability and availability of the system, usually three-tier architecture is adopted in LVS (Linux Virtual Server) clusters illustrated in the following figure 2.1.

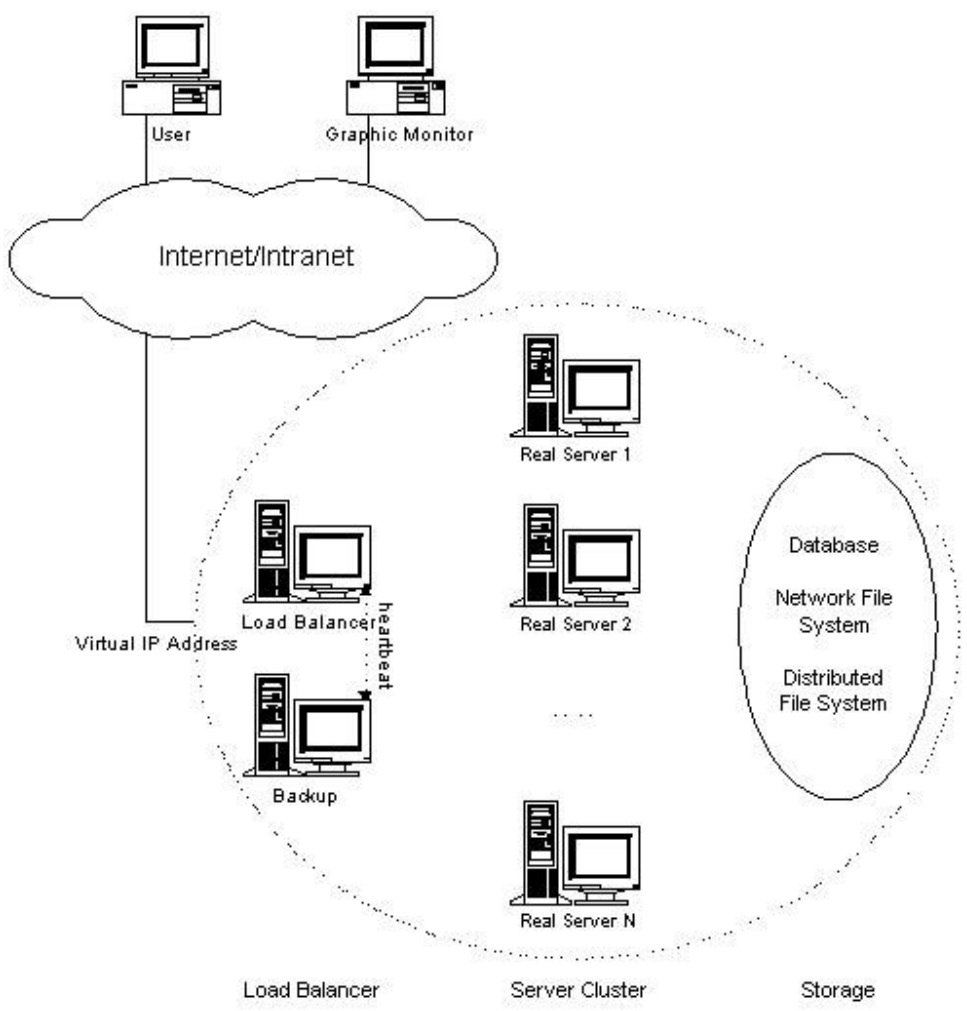

Figure 2.1: Three Tier Architecture of LVS Clusters source: http://www.linuxvirtualserver.org/architecture.html.

Load balancer is the single entry-point of server cluster systems, it can run IPVS (IP Virtual Server) [5] that implements IP load balancing techniques inside the Linux kernel, or KTCPVS stands for Kernel TCP Virtual Server [6]. It implements application-level load balancing inside the Linux kernel, so called Layer-7 switching that implements application-level load balancing inside the Linux kernel [4]. The node number of server cluster can be changed according to the load that system receives. When all the servers are overloaded, more new servers can be added to handle increasing workload. For most Internet services such as web, the requests are usually not highly related, and can be run parallel on different servers. Therefore, as the node number of server cluster increases, the performance of the whole can almost be scaled up linearly [4].

Shared storage can be database systems, network file systems, or distributed file systems. The data that server nodes need to update dynamically should be stored in data based systems, when server nodes read or write data in database systems parallel, database systems can guarantee the consistency of concurrent data access. The static data is usually kept in network file systems such as NFS and CIFS, so that data can be shared by all the server nodes.

## III. RELATED WORKS AND MOTIVATION

Many researchers [12], [13], [14], [15], [16], [17], [18] worked on their Research on Load Balancing of Web-server cluster System in the three tier architecture model. Z. Han and Q. Pan [7] focuses on how to build an LVS load-balancing cluster technology combined with virtualization and shared storage technology to achieve the three-tier architecture of Web server clusters. Y Jiao, W Wang [8] designs and implements load- balancing system of distributed system based web-servers, their design cut costs by reducing time; but does not cut cost by reducing hardware resources, as this system also in the three tier architecture model. C Zheng, J Xia, Q Wang, X Chu [9] design and implement a model to adopt a web-server cluster systems with load balancing algorithms with multiple parameters. Their paper proposes the implementation of real time monitoring status of tasks and dynamic dispatch strategy in web-sever cluster systems [9]. A Krioukov, P Mohan, S Alspaugh and L Keys [10] design a system architecture for web service applications in a standard three-tier architecture model and implement as a power-proportional web cluster. Jiang, Hongbo, et al [11] Present Design, Implementation, and Performance of A Load Balancer for distributing Session Initiation Protocol Server Clusters, and several load balancing algorithms for distributing Session Initiation Protocol (SIP) requests to a cluster of SIP servers in the three tier architecture. Many researchers worked on their Research on Load-Balancing Algorithm for Web Server Clusters [19], [20], [21], [22], [23], these algorithms are being well implemented in different projects in the three tier architecture model.

## IV. PROPOSED MODEL DESIGN AND IMPLEMENTATION DETAILS

For this design model implementation, some open-source software are installed and configured as prototype and tested on the Center for Innovation and Technologies (CIT) Lab, Research center for Science and Technology at DIU. In this experiment, two set of models are conducted to implement high availability Clusters for application load balancing - (1) Web Server Cluster with load balance infrastructure for web servers in three-tier architecture model; (2) Web Server Cluster with load balance infrastructure for web servers in two-tier architecture model.

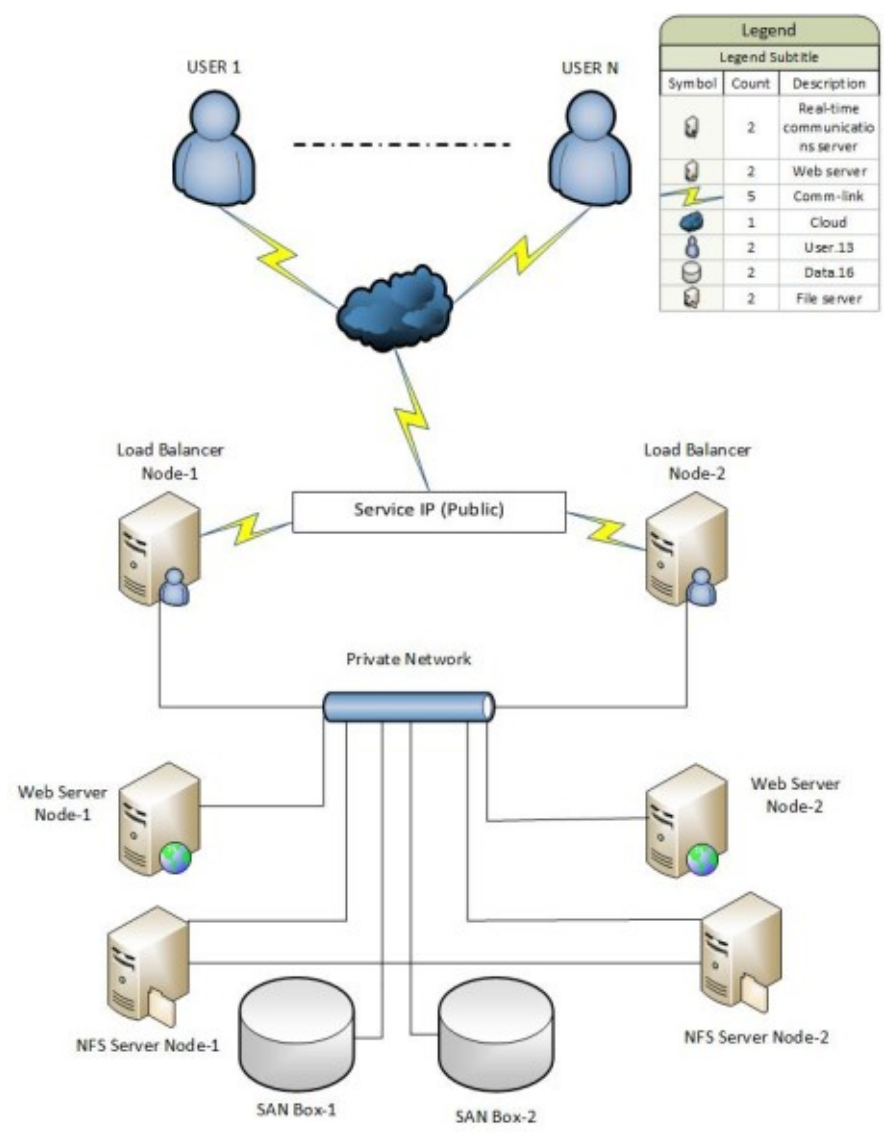

Figure 4.1: Web Server Cluster with Load Balance Infrastructure for web servers in three-tier architecture model.

*A. Three-tier architecture model*

Requirements for hardware resource in the three-tier architectural model of high availability cluster combined with load-balancing technology and shared storage technology for any application load balancing; these four types of nodes are needed – (1) Cluster nodes, (2) Load balancer nodes, (3) SAN Box, and (4) Network File System (NFS) Server Nodes. The typical design of these systems a "High Availability Cluster with Load Balance Infrastructure in three-tier architecture model" describe in the figure 4.1.

In the figure 4.1 real servers and the load balancers are interconnected with high-speed LAN (Private Network). The load balancers dispatch requests to the different servers and make parallel services of the cluster to appear as a virtual service on a single IP address, and request dispatching can use IP load balancing technologies or application-level load balancing technologies. Scalability of the system is achieved by transparently adding or removing nodes in the cluster. High availability is provided by detecting node or daemon failures and reconfiguring the system appropriately. All requests will be processed by Round Robin Algorithm.

*B. Proposed Three-two architecture*

In the two-tier architecture model these three types of nodes are needed – (1) Cluster nodes, (2) Load balancer nodes, and (3) SAN Box. We physically omit Network File System (NFS) Server Nodes and implements within the Load balancer node. The figure 4.2 describes the cost effective proposed two-tier architecture model for High Availability Cluster with Load Balance Infrastructure for web servers.

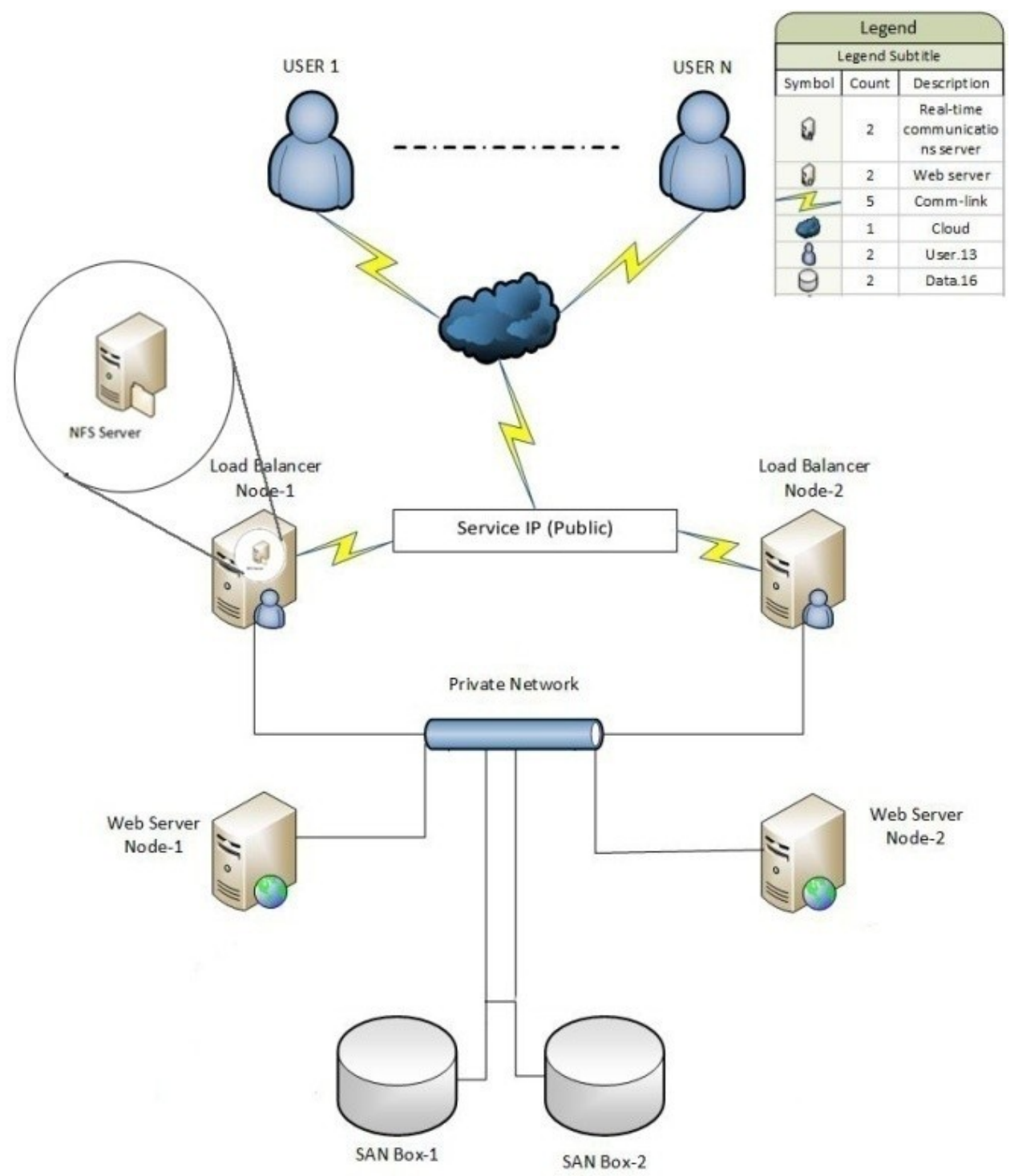

Figure 4.2: Web Server Cluster with Load Balance Infrastructure for web servers in two-tier architecture model.

In this design we used 2 nodes for web server Clusters, 2 nodes for Load balancer and two SAN. For any customized system the modified design can be used up-to 16 nodes for Web Server Clusters, any numbers (n) of nodes for Load balancer. For this system some open-source software are used for configuration and implementation; these are – Red Hat Enterprise Linux (RHEL) 6.4 for load balancer nodes and Cluster nodes; OpenFiler for SAN; Red Hat Cluster Suite (RHCS) for 2 cluster nodes; HAProxy for 2 load balancer nodes; Apache, Network File System (NFS) and PHP.

## V. PERFORMANCE TESTING

To test High Availability Cluster and Load Balancing need some GUI tools and command line tools. For this testing PuTTy, Command Prompt (DOS), cURL, Web Browser are required. The figure 5.1 shows Shared Storage for Load Balancing Node which service (NFS) is configured in Clusters.

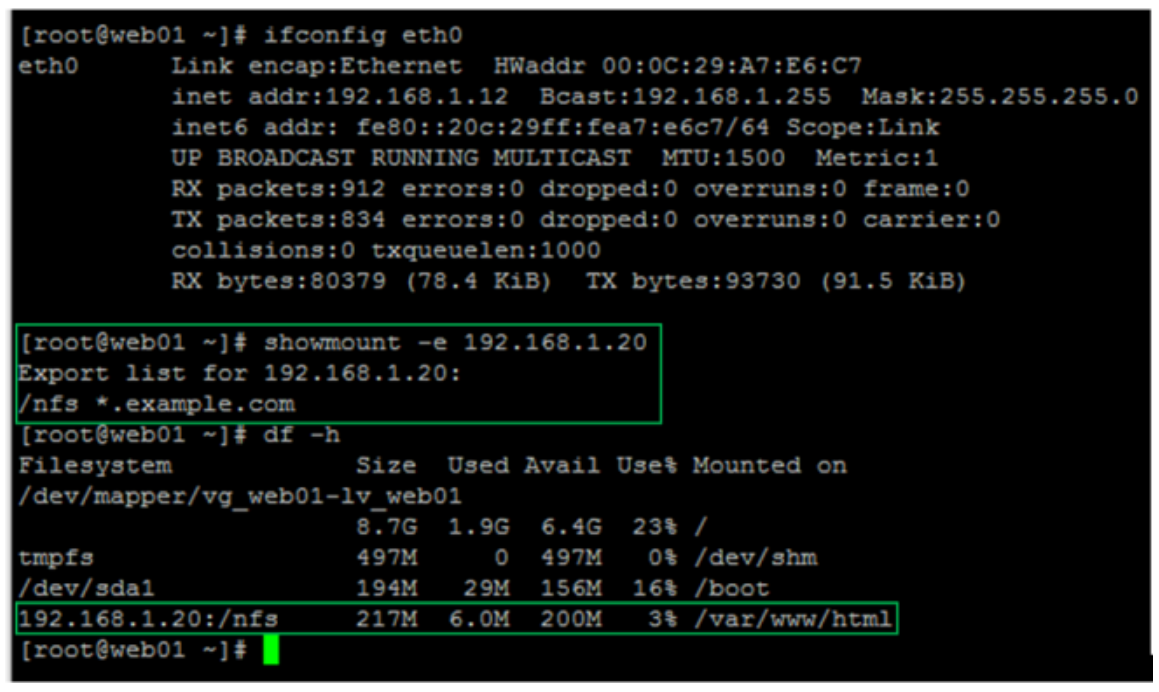

Figure 5.1: Status of Shared Storage for Load Balancing Node which service (NFS) is configured in Cluster

Now we test of High Availability Cluster with these status (in the figure 5.2 and figure 5.3) we can make sure High Availability Cluster is running status is ok. Here service name HAPC and owner hap01.

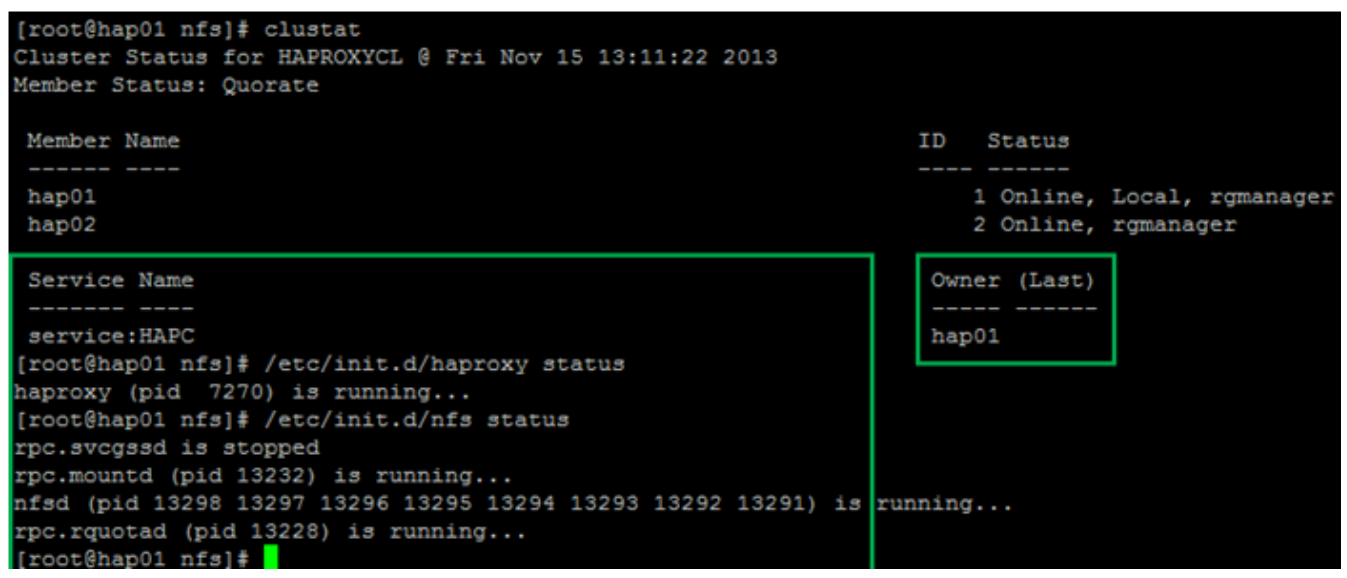

Figure 5.2: HAP01.EXAMPLE.COM Node Status

Figure 5.2 and figure 5.3 both show the status of Cluster service. Currently all Application are running on HAP01.EXAMPLE.COM node and all Application are standby on HAP02.EXAMPLE.COM node. If incase any kind of failure like Network, Hard Disk, Application then all Application will move to the Standby node within very short time. Without any kind of notification to the client the system will overcome failover.

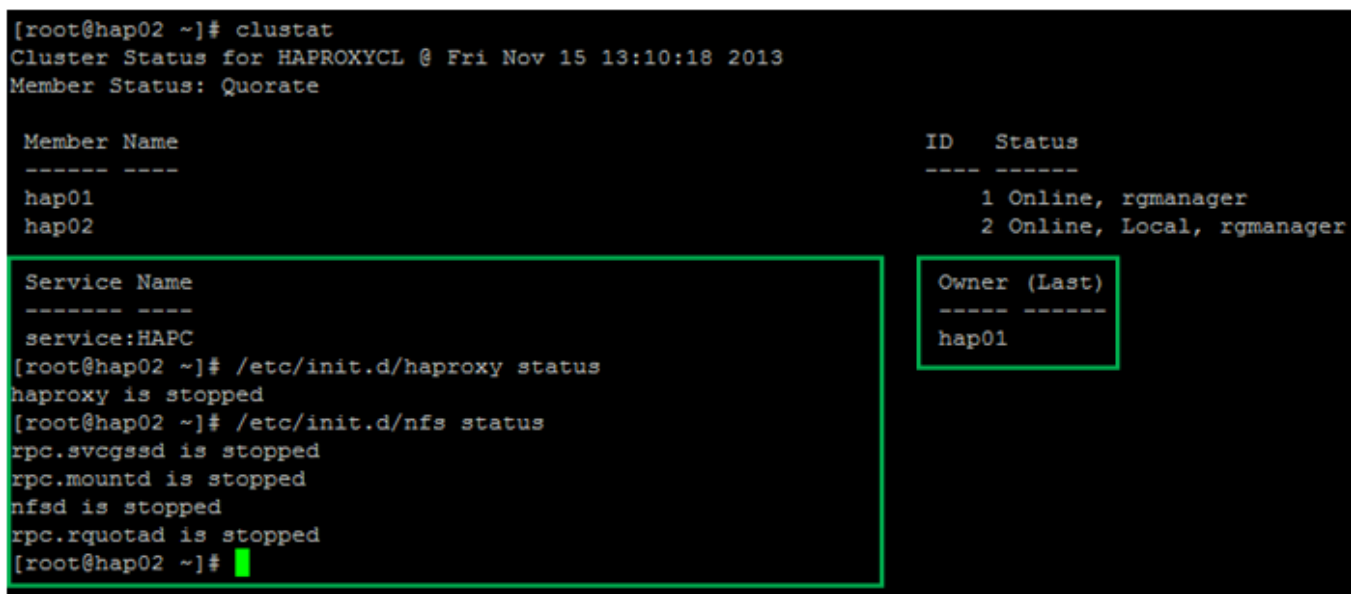

Figure 5.3: HAP02.EXAMPLE.COM Node Status

*Testing of Load Balancing using GUI*

Open any browser and type cluster node Virtual IP in this example here IP is: 192.168.1.20. In production environment this IP should be Public IP.

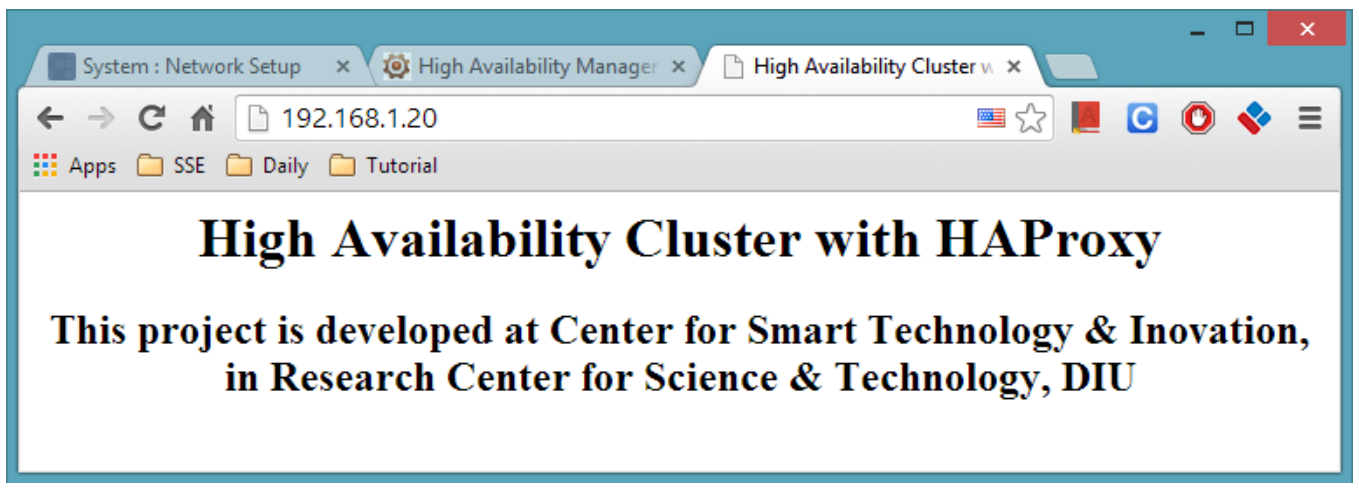

Figure 5.4: Requests serve by Cluster Application Server

In the figure 5.4 shows, node is changed for each request for server with same Virtual IP 192.168.1.20. The changing for node is hidden from the users.

*Testing of Load Balancing using cURL*

First request for server is served by node02 then subsequent by node01 then after node02 and go on serving request in this way (in the figure 5.5). If system has 16 numbers of nodes, then requests are served subsequently 01 number to 16 number nodes then again 01 node. Every request handled by new server (in the figure 5.5). This system managed by an algorithm which is called Round-Robin algorithm or Round-Robin scheduling which is handled request in parallel.

The summarizes of performance test as follows: First Shared Storage for Load Balancing Node for servicing (NFS) is configured in Cluster; after that High Availability Cluster is tested with showing the statuses of Clusters. Load balancing performance is tested using GUI and finally tested by using cURL. Result shows that every request is handled by new server by round-robin scheduling and the system will overcome failover by moving all application to the standby node within very short time.

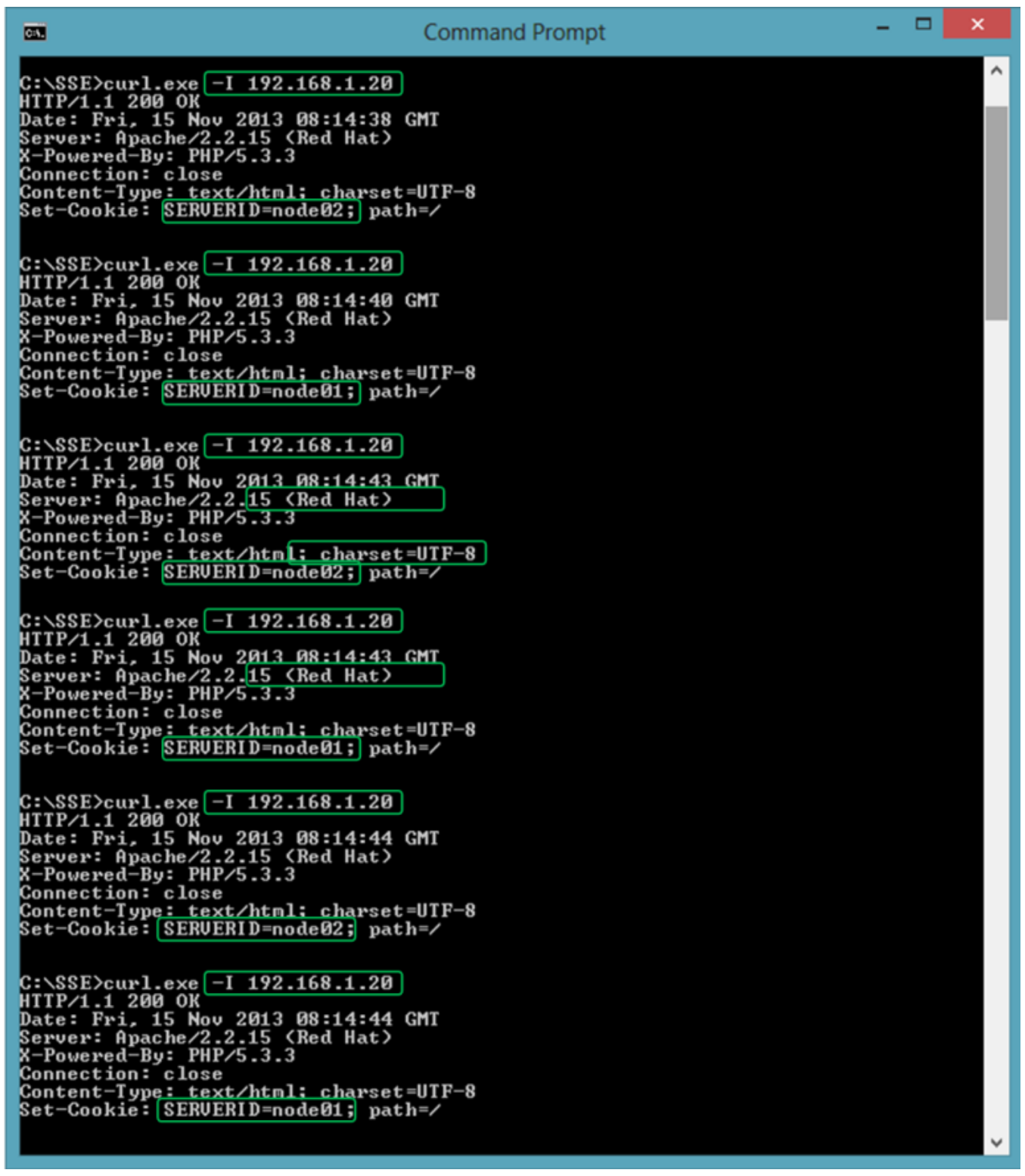

Figure 5.5: Load Balancing Testing using cURL

## VI. Conclusion

This article designs and implements a low cost two-tier architecture model for high availability cluster combined with load-balancing technology and shared storage technology to achieve full facilities of the three-tier architecture for application load balancing e.g. web servers. The paper described design physically omits Network File System (NFS) Server Nodes and implements NFS server functionalities within the Cluster Nodes, through Red Hat Cluster Suite (RHCS) with High Availability (HA) Proxy load balancing technologies - in order to achieve low-cost investment in expensive hardware and computing solutions. For this design implementation, we use 2 nodes for web server Clusters, 2 nodes for Load balancer and two SAN. For any customized system the modified design can be used up-to 16 nodes for Web Server Clusters, any numbers (n) of nodes for Load balancer. For this system some open-source software are used for configuration and implementation; these are – Red Hat Enterprise Linux (RHEL) 6.4 for load balancer nodes and Cluster nodes; OpenFiler for SAN; Red Hat Cluster Suite (RHCS) for 2 cluster nodes; HAProxy for 2 load balancer nodes; Apache, Network File System (NFS) and PHP.

This system can provide continuous service though any system components fail uncertainly such as network system, storage and application. This system can overcome in the cases of node failover, network failover, storage limitation and distributions load as like as the all the facilities of three-tier architecture model for high availability cluster for application load balancing with limited physical resources for low cost implementation.